\documentclass[10pt]{article}

\usepackage{graphicx}
\usepackage{float}
\usepackage[many]{tcolorbox}
\usepackage{booktabs}
\usepackage{hyperref}
\usepackage{upgreek}
\usepackage{cite}
\usepackage{amsmath}
\usepackage{amsthm}
\usepackage{amsfonts}
\usepackage[textwidth=16cm]{geometry}
\usepackage{caption}
\newcommand{\fon}[1]{\fontfamily{#1}\selectfont}
\NewTColorBox{NewBox}{ s O{!htbp} }{%
  floatplacement={#2},
  IfBooleanTF={#1}{float*,width=\textwidth}{float},
  colframe=black,colback=white,
  fonttitle=\fon{phv},
  }

\title{A protocol for dynamic model calibration}

\author{
{\small
Alejandro F. Villaverde\,$^{\text{1}}$, Dilan Pathirana\,$^{\text{2}}$, Fabian Fr\"ohlich\,$^{\text{3,4,5}}$, Jan Hasenauer\,$^{\text{2,3,4},\ast}$ and Julio R. Banga\,$^{\text{6},\ast}$ } \\
{\small
$^{\text{1}}$Universidade de Vigo, Department of Systems Engineering \& Control, Vigo 36310, Galicia, Spain } \\
{\small$^{\text{2}}$Faculty of Mathematics and Natural Sciences, University of Bonn, Bonn 53115, Germany } \\
{\small$^{\text{3}}$Institute of Computational Biology, Helmholtz Zentrum M\"unchen, Neuherberg 85764, Germany } \\
{\small$^{\text{4}}$Center for Mathematics, Technische Universit\"at M\"unchen, Garching 85748, Germany } \\
{\small$^{\text{5}}$Harvard Medical School, Cambridge, MA 02115, USA } \\
{\small$^{\text{6}}$Bioprocess Engineering Group, IIM-CSIC, Vigo 36208, Galicia, Spain } \\
{\small$^\ast$To whom correspondence should be addressed: jan.hasenauer@uni-bonn.de, j.r.banga@csic.es.}
}

\date{}

\begin{document}

\maketitle 

\abstract{
Ordinary differential equation models are nowadays widely used for the mechanistic description of biological processes and their temporal evolution. These models typically have many unknown and non-measurable parameters, which have to be determined by fitting the model to experimental data. In order to perform this task, known as parameter estimation or model calibration, the modeller faces challenges such as poor parameter identifiability, lack of sufficiently informative experimental data, and the existence of local minima in the objective function landscape. These issues tend to worsen with larger model sizes, increasing the computational complexity and the number of unknown parameters. An incorrectly calibrated model is problematic because it may result in inaccurate predictions and misleading conclusions. For non-expert users, there are a large number of potential pitfalls. Here, we provide a protocol that guides the user through all the steps involved in the calibration of dynamic models. We illustrate the methodology with two models, and provide all the code required to reproduce the results and perform the same analysis on new models. Our protocol provides practitioners and researchers in biological modelling with a one-stop guide that is at the same time compact and sufficiently comprehensive to cover all aspects of the problem.\\
\textbf{Key words: }
systems biology; dynamic modelling; parameter estimation; identification; identifiability; optimisation.
}

\section*{Introduction}

The use of dynamic models has become common practice in the life sciences. Mathematical modeling provides a rigorous, compact way of encapsulating the available knowledge about a biological process. Perhaps more importantly, it is also a tool for understanding, analysing, and predicting the behaviour of a complex system under conditions for which no experimental data are available. To these ends, it is particularly important that the model has been developed with that specific purpose in mind. 

In biomedicine, dynamic models are used for basic research as well as for medical applications. On one hand, dynamic models facilitate an understanding of biological processes, e.g. by identifying from a list of alternative mechanisms the most plausible one \cite{kuepfer2007ensemble}. On the other hand, dynamic models with sufficient mechanistic detail can be used to make predictions, including the selection of drug targets \cite{SchoeberlPac2009}, and the outcome of individual and combination treatments \cite{frohlich2018efficient,henriques2017seldom}. 
In bio- and process engineering, dynamic models are used to design and optimise biotechnological processes. Here, models are, for instance, used to find the genetic and regulatory modifications that enhance the production of a target metabolite while enforcing constraints on certain metabolite levels \cite{song2013modeling,almquist2014kinetic,villaverde2016metabolic,briat2018perfect}. 
In synthetic biology, dynamic models guide the design of artificial biological circuits where fine-tuned expression levels are necessary to ensure the correct functioning of regulatory elements \cite{karamasioti2017computational,hsiao2018control,steel2018design,tomazou2018computational}. Beyond these topics, there is a broad spectrum of additional research areas.

The choice of model type and complexity depends on which biological question(s) it should address. Once this has been decided, the relevant biological knowledge is collected, e.g. from databases such as KEGG \cite{kanehisa2000kegg}, STRING \cite{szklarczyk2016string}, and REACTOME \cite{fabregat2017reactome}, or from the literature. Furthermore, already available models can be used, e.g. from JWS Online \cite{olivier2004web} or Biomodels \cite{BioModels2006}, and information about kinetic parameters might be extracted, e.g. from BRENDA \cite{chang2014brenda} or Sabio-RK \cite{wittig2012sabio}. This information is used to determine the biological species and biochemical reactions that are relevant to the process. In combination with assumptions about reaction kinetics -- e.g. mass action or Michaelis-Menten -- these elements allow the construction of a tailored mathematical model, which will usually have nonlinear dynamics and uncertainties associated to its structure and parameter values \cite{vanRiel:2006}. The model can be specified in a standard format such as SBML, to take advantage of the ecosystem of tools that already support standard format \cite{hucka:2003}. 

The advent of high-throughput experimental techniques and the ever-growing availability of computational resources have led to the development of increasingly larger models. Common models possess tens of state variables and tens to a few hundreds of parameters (see \cite{villaverde2018benchmarking,hass2018benchmark}). Large models can even possess thousands of state variables and parameters \cite{frohlich2018efficient}. Dynamic models need to be calibrated, i.e. their unknown parameters have to be estimated from experimental data. In model calibration, the mismatch between simulated model output and experimental data is minimised to find the best parameter values \cite{jaqaman-danuser:2006,ashyraliyev2009systems,Geier2012,Raue2013,gadkar2016six}. Model calibration is a process composed of a sequence of steps, which usually need to be iterated \cite{balsa-alonso-banga:2010} until a satisfactory result is found. It may be seen as part of a more general problem sometimes called reverse engineering \cite{villaverde2014reverse} or (nonlinear) systems identification \cite{schoukens2019nonlinear}.

In this work, we consider the calibration of ordinary differential equation (ODE) models. ODE models are widely used to describe biological processes, and their calibration has been discussed in protocols for different classes of processes, including gene regulatory circuits \cite{seaton2017ode}, signalling networks \cite{Geier2012}, biocatalytic reactions \cite{eisenkolb2019modeling}, wastewater treatment \cite{mannina2011practical,zhu2015novel}, food processing \cite{vilas2018toward}, biomolecular systems \cite{tuza2019systematic}, and cardiac electrophysiology models \cite{whittaker2020calibration}. Yet, these protocols focus on individual aspects of the calibration process (relevant for the sub-discipline) and/or lack illustration examples and codes that can be reused.
The papers \cite{mannina2011practical} and \cite{zhu2015novel} focus on parameter subset selection via sensitivity and correlation analysis, and on subsequent model optimisation. The works of \cite{seaton2017ode}, \cite{vilas2018toward} and \cite{eisenkolb2019modeling} consider only low-dimensional models and do not provide in-depth discussion of scalability. The paper \cite{Geier2012} neither covers structural identifiability analysis nor experimental design, and describes a prediction uncertainty approach with limited applicability. The works of \cite{tuza2019systematic}, \cite{whittaker2020calibration} and \cite{eisenkolb2019modeling} discuss most aspects of the calibration process, but do not provide a step-by-step illustration with an example model and codes. The work of \cite{steiert2019recipes} is tailored to users of the MATLAB software toolbox Data2Dynamics \cite{raue2015data2dynamics}.

\begin{figure*}[!tpb]
\centerline{\includegraphics[width=0.9\linewidth]{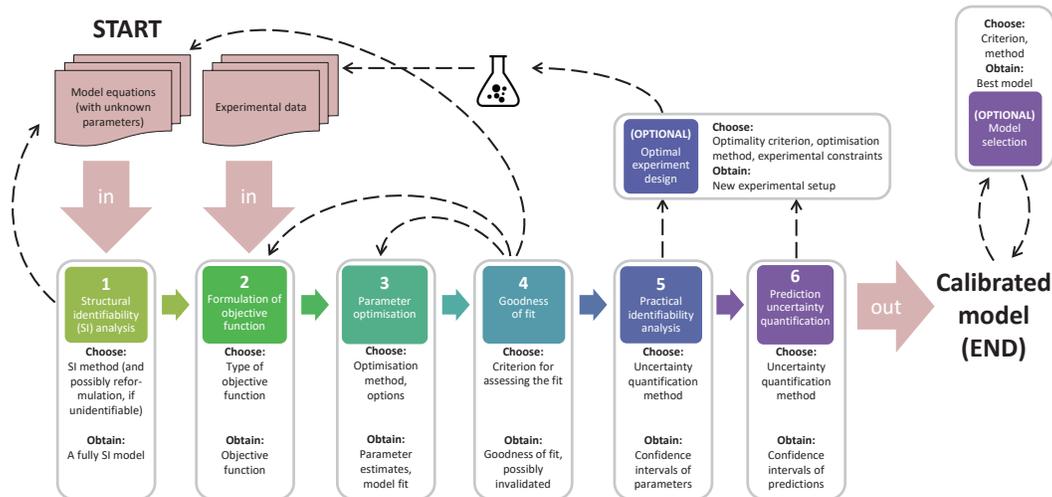}}
\caption{Block diagram of the model calibration process presented in this protocol.}\label{fig:01}
\end{figure*}

This protocol aims to provide a comprehensive description of the steps of the calibration process, which integrates recent advances. An outline of the procedure is depicted in Figure~\ref{fig:01}.
The article is structured as follows. First we describe the requirements for running the calibration protocol. Then, we describe the individual steps of the protocol. 
The theoretical background for each step, along with a brief review of available methodologies, is provided in 
boxes.
After some troubleshooting advice, we illustrate the application of the protocol for two case studies. 
For the sake of clarity, only a concise summary of the application results is reported in the main text of this manuscript; complete details are given in the supplementary information.
To ensure the reproducibility of the results, we provide computational implementations used for the application of the protocol steps to the case studies in the form of MATLAB live scripts, Dockerfiles, and Python-based Jupyter notebooks.

\section*{Materials}

This section describes the inputs and equipment required to run the protocol.

\paragraph{Hardware:} a standard personal computer, or a computer cluster.
For demonstrating the application of the protocol, in the present work we have performed Step 1 on a standard laptop with a 2.40 GHz processor and 8 GB RAM. Optimisation, likelihood profiling, and sampling were performed on a laptop with an Intel Core i7-10610U CPU (eight 1.80GHz cores) and 32 GB RAM, with a total runtime of up to 2 days, per model.

\paragraph{Software:}     
a software environment with numerical computation and visualisation capabilities, along with specialised toolboxes that facilitate performing specific protocol steps. 
Table \ref{Tab:materials} lists the software resources used in this work.

\paragraph{Model:}   
a dynamic model described by nonlinear ODEs of the following form:
\begin{equation}\label{eq:model}
	\begin{aligned}
	\dot{x} &= f\left(x,\theta,t \right), \ x(t_0) = x_0(\theta), \\
	y       &= g(x,\theta,t),
	\end{aligned}
\end{equation}
in which $x(t)\in\mathbb{R}^{n_x}$ is the state vector at time $t$ with initial conditions $x_0(\theta)$, $y(t)\in\mathbb{R}^{n_y}$ is the output (i.e. observables) vector at time $t$, $f$ and $g$ are possibly nonlinear functions, and $\theta\in\mathbb{R}^{n_\theta}$ is the vector of unknown parameters. 

In this work we used a carotenoid pathway in \textit{Arabidopsis thaliana} \cite{bruno2016enzymatic}, and an EGF-dependent Akt pathway of the PC12 cell line \cite{fujita2010decoupling}, taken from the PEtab benchmark collection \cite{hass2018benchmark} available at \url{https://github.com/Benchmarking-Initiative/Benchmark-Models-PEtab}. An illustration of both models is provided in panels A of Fig. \ref{fig:bruno} and Fig. \ref{fig:fujita}.

\paragraph{Data:} a set of time-resolved measurements of the model outputs.
In the present work, data was taken from the aforementioned PEtab benchmark collection.

\begin{table*}[!thb]
\centering
\resizebox{\textwidth}{!}{
\begin{tabular}{l l l l l l}
\toprule 
Name          & Type        & Steps & Reference              & Website                         & Environment\\
\midrule
MATLAB        & environment & all   &                        & \url{http://www.mathworks.com} &    \\
Python        & environment & all   &                        & \url{https://www.python.org}  &    \\
\midrule
SBML          & model format& input & \cite{hucka:2003}      & \url{http://www.sbml.org}             & MATLAB, Python \\
PEtab         & data format & input & \cite{schmiester2021petab} & \url{https://github.com/PEtab-dev/PEtab} & Python \\
\midrule
STRIKE-GOLDD  & tool (SI analysis)     & 1     & \cite{villaverde2019full} & \url{https://github.com/afvillaverde/strike-goldd} & MATLAB \\
AMICI         & tool (simulation)   & 2 & \cite{frohlich2017scalable} & \url{https://github.com/AMICI-dev/AMICI} & Python    \\
pyPESTO       & tool (various steps) & 3, 5, 6  & \cite{stapor2017pesto, yannik_schalte_2021_4738579} & \url{https://github.com/ICB-DCM/pyPESTO} & Python  \\
Fides         & tool (param. optimisation) & 3, 5 & \cite{Froehlich2021fides} & \url{https://github.com/fides-dev/fides} & Python \\
SciPy         & tool (various steps) & 3, 5     & \cite{2020SciPy-NMeth}    & \url{https://www.scipy.org} & Python \\
Data2Dynamics & tool (various steps) & 3, 5, 6, (O) & \cite{raue2015data2dynamics} & \url{http://www.data2dynamics.org}   & MATLAB \\
\midrule
\end{tabular}
}
\caption{\label{Tab:materials}Software resources for dynamic model calibration used in this work.}
\end{table*}

\section*{Procedure}

The protocol consists of six main steps, numbered 1--6, which consist of sub-steps. Furthermore, we describe two optional steps. The workflow is depicted in Fig.~\ref{fig:01} and described in the following paragraphs. 

\subsection*{STEP 1: Structural identifiability analysis}\label{sec:si} 

\begin{figure*}[!ht]
\centering
\centerline{\includegraphics[width=0.9\linewidth]{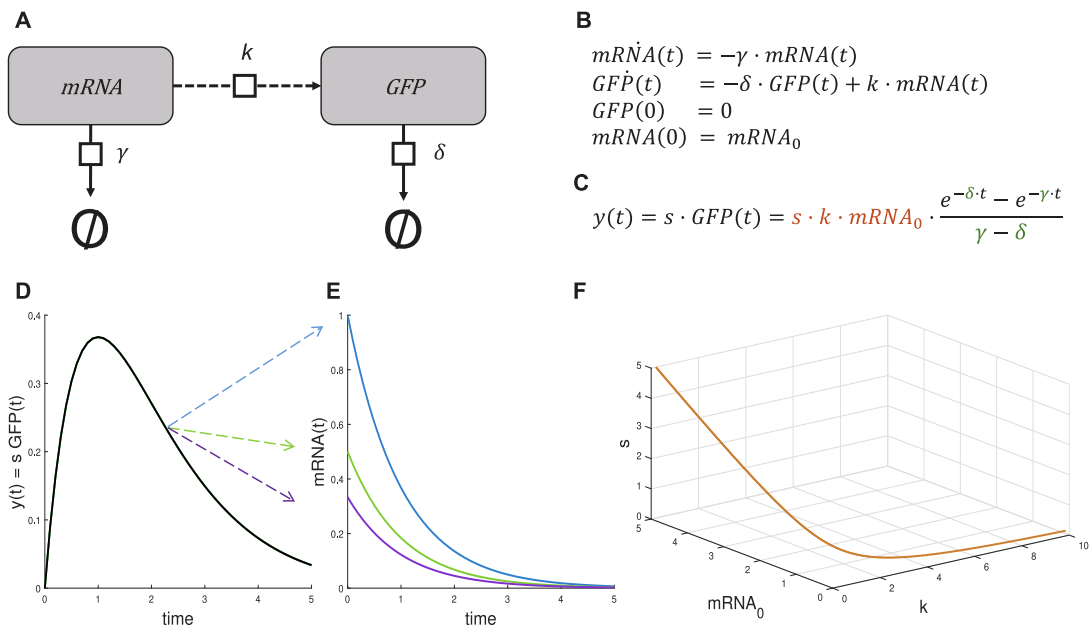}}\caption{\textbf{Structural identifiability analysis.}
\textbf{(A)} Diagram of a simplified model of mRNA translation considering only the process in the cytosol. The model captures the translation of mRNA and the degradation of mRNA and protein. 
(B) Mathematical formulation (ODEs) of mRNA translation dynamics \cite{ballnus2018bayesian} involving two states, mRNA and GFP. 
(C) The model output is the fluorescence intensity, which is proportional to the GFP level. The model has five unknown parameters: the initial condition of the unmeasured state (mRNA$_0$), three kinetic parameters ($\gamma, \delta, k$), and an output scaling parameter ($s$). Given its simplicity, it is possible to calculate the output time-course analytically (here shown for $\gamma \neq \delta$). The resulting function contains the product of three parameters ($s\cdot k\cdot mRNA_0$), which is shown in orange, and an expression involving $\delta$ and $\gamma$, which are shown in green. The latter expression is symmetrical with respect to $\delta$ and $\gamma$: their values can be exchanged without changing the result. Thus, these two parameters are not structurally globally identifiable, but only locally identifiable with two possible solutions. Furthermore, the product ($s\cdot k\cdot mRNA_0$) allows for an infinite number of parameter combinations; the three involved parameters are structurally non-identifiable. 
(D) Illustration of structural non-identifiability: the time-course of the model output is identical for an infinite number of parameter vectors. 
(E) Illustration of unobservability caused by non-identifiability. For illustration purposes, three different parameter vectors are shown, all of which produce the same model output. Each of them yields a different simulation of the mRNA time-course; thus, this state cannot be determined. (F) Illustration of the correlations between the non-identifiable parameters. The line indicates parameter combinations for which the time-dependent output is identical.
}\label{fig:si}
\end{figure*}

Structural identifiability is analysed to assess whether the values of all unknown parameters can be determined from perfect continuous-time and noise-free measurements of the observables under the given set of experimental conditions \cite{walter-pronzato:97,distefano2015dynamic}. 
Structural non-identifiabilities imply that there are several model parameterizations, e.g. due to symmetries or redundancies in the model structure, which yield exactly the same observables. An overview of the available methodologies for structural identifiability analysis is provided in Box 1.  
Fig.~\ref{fig:si} illustrates possible sources of structural non-identifiability and the related issues.
The structural identifiability analysis can be complemented by observability analysis, which determines if the trajectory of the model state can be uniquely determined from the observables.

\begin{NewBox}*
\textbf{Box 1. Methods for STEP 1: Structural identifiability analysis}\\
\setlength\parindent{12pt}
Structural identifiability (SI) can be analysed using a broad spectrum of methods exploiting, e.g., Taylor and generating series, differential algebra, differential geometry, and probabilistic numerics \cite{miao-xia-perelson-wu:2011,chis-banga-balsa-canto-plos:2011,villaverde2019observability}. In essence, these methods aim to assess whether the mapping from parameters to observables is invertible for almost all points in parameter space.

When choosing a method, the trade-off between generality and applicability must be taken into account. Most available methods are tailored to rational models, i.e. $f$ and $g$ can be expressed as fractions of polynomials. Structural local identifiability of rational models can be assessed efficiently using e.g. the exact arithmetic rank method in \cite{sedoglavic2002probabilistic,anguelova2012efficient}. 
Structural global identifiability analysis requires more computationally expensive techniques that do not scale well, hence, this approach can only be applied to models with tens of parameters and state variables. For non-rational models, a higher-dimensional polynomial or rational model can be formulated with an identical input-output map \cite{ohtsuka2005model}. This immersion shifts the non-rational relations from the vector field to constraints on the initial conditions. These constraints can be relaxed to apply methods for rational models; however, for the relaxed problem, results about non-identifiability may not be conclusive \cite{chatzis2015observability}.

In the present work we used STRIKE-GOLDD to assess structural local identifiability and observability \cite{villaverde2019full}. Tools for structural global identifiability analysis include GenSSI2 \cite{ligon2017genssi}, SIAN \cite{hong2019sian}, COMBOS \cite{meshkat2014z}, and DAISY \cite{saccomani2019new}.

\end{NewBox}

The first step in the protocol is thus:

\subsubsection*{STEP 1.1} 
Analyse the structural identifiability of the model with one of the methods described in Box 1.

If all parameters are structurally identifiable and all state variables are observable, we continue with Step~2.1. Otherwise, we recommend to determine the source of the structural non-identifiability as an intermediate step (1.2). Ideally, the parametric form of the non-identifiable manifold (i.e. the set of parameters that yield identical observables) is determined. Some tools offer this functionality or at least provide hints.

\subsubsection*{STEP 1.2}
If parameters are structurally non-identifiable or state variables unobservable, use knowledge about the structure of the non-identifiable manifold to 
\begin{itemize}
    \item reformulate the model by merging the non-identifiable parameters into identifiable combinations, OR
    \item fix the non-identifiable parameters to reasonable values.
\end{itemize}
In both cases, the information about the non-identifiability needs to be retained to later perform a proper analysis of the prediction uncertainties. If this point is not taken into account, the obtained results are only valid for the reformulated model, but not for the original one -- a fact that is often disregarded.

An alternative to the reformulation of the model or the fixing of parameters is to plan additional experiments, if possible. These can be experiments with new experimental conditions, new observables, or both (keeping experimental constraints in mind). The additional information should be recorded such that more, ideally all, parameters are structurally identifiable. 

\subsection*{STEP 2: Formulation of objective function}

The objective function measuring the mismatch of simulated model observables and measurement data is defined. The choice of the objective function depends on the characteristics of the measurement technique and accounts for knowledge about its accuracy. Possible choices are discussed in Box 2.

\subsubsection*{STEP 2.1}
Construct an objective function.

\begin{NewBox}*

\textbf{Box 2. Theory for STEP 2: Formulation of objective function}\\
\setlength\parindent{12pt}
The objective function encodes the characteristics of the measurement process and potential prior knowledge. It can be composed of up to two parts:
\begin{itemize}
    \item the likelihood function $p(\mathcal{D}|\theta)$ provides the likelihood of measuring the dataset $\mathcal{D}$ given the model parameters $\theta$, and
    \item the prior distribution $p(\theta)$ encodes additional belief.
\end{itemize}

Frequentist approaches only use the likelihood function, and the common choice of the objective function is the negative log-likelihood function, $J(\theta) = - \log p(\mathcal{D}|\theta)$. Bayesian approaches use the posterior instead of only the likelihood, hence also consider the prior distribution, yielding the unnormalised negative log-posterior $J(\theta) = - \log p(\mathcal{D}|\theta) - \log p(\theta)$ (which disregards the marginal probability). In contrast to the likelihood function and the posterior distribution, the logarithmic transformations are numerically easier to evaluate and allow for faster optimiser convergence~\cite{hass2018benchmark}.

Under the assumption of independent measurements and normally distributed noise, the negative log-likelihood function is
\begin{equation}\label{eq:llk}
	J(\theta) = \frac{1}{2} \sum^{n_y}_{j=1} \sum^{n_t}_{k=1} 
	\left[
	\log
	\left( 2\pi \sigma^2_j(t_k) \right) 
	+
	\left(
	\frac{y^m_j(t_k) - y_j(t_k,\theta)} {\sigma_j(t_k)}\right)^2
	\right],
\end{equation}
in which $y^m_{j}(t_k)$ is the measured value and $y_{j}(t_k,\theta)$ is the simulation result of the $j$-th observable, at time point $t_k$. The corresponding standard deviation of the measurement noise is denoted by $\sigma_{j}(t_k)$, and can be known (e.g. determined from multiple measurement replicates) or a function of the parameters. \footnote{In many applications the measurements are repeated to obtain biological or at least technical replicates. In this case, one can either (i) use all replicates as measurements and set the noise level to the standard deviation, or (ii) use the mean of the replicates as measurement and set the noise level to the standard error of mean. A combination of (i) and (ii) is statistically not meaningful.} In the case of known standard deviations, the summands $\log( 2\pi \sigma^2_j(t_k))$ are independent of the parameter vector, and can be disregarded for parameter optimisation and uncertainty analysis. This yields the weighted least squares objective function:
\begin{equation}\label{eq:statement}
	J(\theta) = \sum^{n_y}_{j=1} \sum^{n_t}_{k=1} 
	w_j(t_k) \left( y^m_j(t_k) - y_j(t_k,\theta)\right)^2
\end{equation}
with $w_j(t_k) = \sigma^{-2}_j(t_k)$. Sometimes this objective function is also applied without proper statistical motivation, e.g. without linking the weights to the noise levels, which does not yield a proper frequentist formulation of the calibration problem.

The likelihood function is based on the assumed or observed probability distribution of the experimental error. Normal and log-normal distributions are common choices \cite{hengl-kreutz-timmer-maiwald:07}, but in a recent study Laplace distributions have also been used to achieve robustness against outliers \cite{MaierLoo2017}. For count measurement, distributions such as the binomial and negative binomial distributions appear to be suited. Furthermore, the consideration of the correlation of measurement errors might be necessary. 
An alternative to statistically motivated prior distributions $p(\theta)$ are more mathematically motivated regularisation functions. In general, regularisation is used to tackle the problem that models are often over-parameterised. In this case, the calibration problem is ill-posed and small perturbations in the data can result in very different parameter estimates \cite{hadamard1902problemes}. Regularisation techniques address this problem by penalizing undesirable parameter choices. Tikhonov regularisation penalises large parameter values, and pushes estimates towards zero. Mathematically, it is identical to a normally distributed prior $p(\theta)$ with mean zero. Alternative regularisation schemes include \cite{lopez2015nonlinear} parameter subset estimation, truncated singular value decomposition, principal component analysis, and Bregman regularisation.

The objective functions encountered in model calibration are mostly nonlinear and non-convex, and possess multiple optima. Furthermore, there are frequently flat regions (in particular in the presence of non-identifiabilities), and rims (e.g. at bifurcation points).

\end{NewBox}

\subsection*{STEP 3: Parameter optimisation}

\begin{figure*}
\centering
\includegraphics[width=0.9\linewidth]{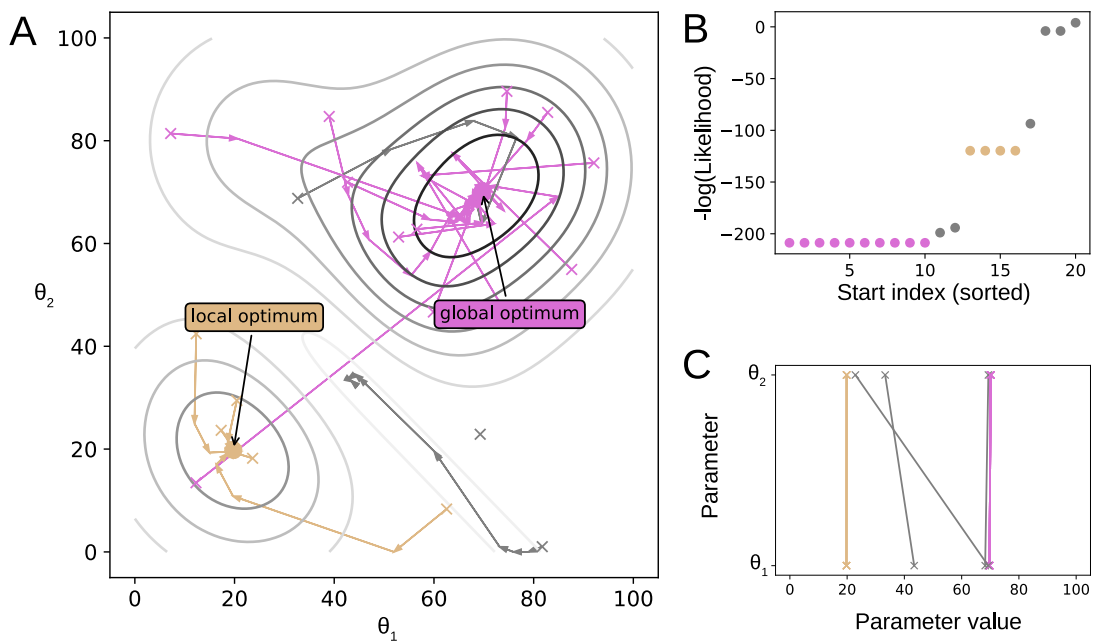}
\caption{\textbf{Parameter optimisation.} (A) Multi-start local optimisation involves many local optimisations that are distributed within the parameter space. In systems with multiple optima, many starts may be required to find the global optimum. Trajectories are indicated by arrows, with their initial points indicated with ``$\times$''. The contour plot shows the negative log-likelihood, with darker contours indicating lesser (better) values. In all subfigures, the colours pink (global) and brown (local) are used to indicate results that correspond to a particular optimum, and parameters are labelled as $\uptheta$ with an index as the subscript. This subfigure is for illustration purposes only, as it is generally infeasible to produce. (B) Convergence of starts towards an optimum can be assessed with a waterfall plot, where the existence of (multiple) plateaus indicates optimiser convergence. If plateau(s) are not seen, possible solutions include: additional starts; alternative initial points; or alternative global optimisation methods. (C) A parallel coordinates plot can be used to assess whether parameters are well-determined. Here, lines belonging to a single optimum overlap, indicating that the parameters that have converged to the corresponding optimum are well-determined.}
\label{fig:concept_optimise}
\end{figure*}

Parameter estimates are obtained by minimising the objective function. To this end, numerical optimisation methods suited for nonlinear problems with local minima should be employed. Available methodologies and practical tips for their application are discussed in Box 3, and key aspects are illustrated in Fig. \ref{fig:concept_optimise}.

\subsubsection*{STEP 3.1}
Launch multiple runs of local, global, or hybrid optimisation algorithms. The number of runs required is model-dependent. For an initial optimisation we recommend at least 50 runs with purely local searches, or at least 10 runs with global or hybrid searches.

Accurate gradient computation is required for gradient-based optimisation. Before optimisation, check that the gradients appear correct by evaluating the gradient at a point, and then compare this with forward, backward, and central finite difference approximations of the gradient that are evaluated with different step sizes. Such a gradient check is a common, possibly optional, feature of tools that provide gradient-based optimisation.

\subsubsection*{STEP 3.2}
Evaluate the reproducibility of the fitting results by comparing the optimal objective function values achieved by different runs. The optimal objective function values should be robustly reproducible, meaning that a substantial number of runs (rule-of-thumb: 5) should find it. If this is not the case, repeat Step 3.1 with a larger number of runs.
Note that the difference between runs that is considered negligible should be statistically motivated. For the use of log-likelihood and log-posterior this corresponds to an absolute difference, not a relative one \cite{hass2018benchmark}.

\begin{NewBox}*

\textbf{Box 3. Methods for STEP 3: Parameter optimisation.}\\
\setlength\parindent{12pt}
Objective functions encountered in systems biology are usually non-convex and multi-modal. Hence, global optimisation methods are required to robustly determine the optimal solution. Common choices are (i) a multi-start strategy that performs local searches from different starting points \cite{Raue2013}, and (ii) a hybrid methodology combining a metaheuristic algorithm with local searches \cite{villaverde2018benchmarking}.
Independently of the employed method, it is advisable to run the optimisation method several times \cite{hross2016analysis}.

Independently of the setting (multi-start or hybrid) in which the local searches are performed, it is recommended to use a \textit{gradient-based} local method, which exploits the knowledge about the gradient of the objective function to drive the search \cite{villaverde2018benchmarking}. 
Gradient computation using \textit{adjoint sensitivities} has been shown to outperform forward sensitivities and finite differences for medium- and large-scale models \cite{frohlich2017scalable}. Finite differences are least reliable, as an appropriate choice of the step-size is problematic and depends on the parameters. Furthermore, for most parameters it is beneficial to estimate them on a logarithmic scale \cite{Raue2013,kreutz2016new,hass2018benchmark}. 

Many optimisation methods can exploit parallel infrastructure, allowing for a reduction in computation times if a multi-core computer or cluster is available \cite{raue2015data2dynamics,penas2017parameter}.

\end{NewBox}

\subsection*{STEP 4: Goodness of fit}
The quality of the fitted model should be assessed by visual inspection. 
It is also possible to use quantitative metrics for this purpose.
Details are provided in Box~4.

\begin{NewBox}*

\textbf{Box 4. Methods for STEP 4: Assessment of the goodness of fit.}\\
\setlength\parindent{12pt}
The root-mean-square error (RMSE) between simulated and measured observables, i.e. the square root of the mean squared error, provides a quantitative metric of the goodness of fit. The normalised root mean squared error (NRMSE) is obtained by dividing the RMSE by the range of the measurements, and it is usually more useful since it enables a direct comparison between different observables and/or estimation methods. The NRMSE has the additional advantage of being independent of the noise in measurements and the number of data points used for the fitting. These and other metrics are further discussed in \cite{Li2017}.

Complementary to this, the achieved objective function value can be compared with the expected objective function value. The distribution of expected objective function values for the data generating model with the true parameters can be constructed from the knowledge of the model and the measurement setup. For normally distributed measurement noise with known standard deviation, the sum of squared residuals follows a chi-squared distribution with $n_{\mathcal{D}}$ (number of data points) degrees of freedom. However, while the distribution for the true parameter is analytically tractable, for the estimated parameter this is not the case. To assess whether the achieved objective function value is plausible, approximations can be employed. For linear regression problems it is known that the sum of squared residuals at the optimal parameters follows a chi-squared distribution with $n_{\mathcal{D}} - n_\theta$ degrees of freedom, where $n_\theta$ is the number of estimated parameters. This is often also used as an approximation for ODE models \cite{Geier2012}, but for certain models (e.g. those with oscillatory dynamics) this can be off. An alternative is the use of bootstrapping procedures, in which a problem-specific distribution is constructed  \cite{efron1986bootstrap}.

If the achieved objective function value is much larger than most values expected under the distribution, this can indicate underfitting. This implies that either the optimisation was not successful or that the model is inappropriate. If the achieved objective function value is much smaller than expected, this is a sign of overfitting, meaning that not only the signal in the measurement data but the noise is described. Under- and overfitting are possible causes of wrong model predictions and should be avoided.

A way of controlling for overfitting is to perform cross-validation. To this end, a subset of the data must be left out of the optimisation in STEP 3. Afterwards, the calibrated model is used for predicting this data subset. Overfitting appears if the model achieves a good fit in the optimisation, but then fails to generalise to observables or experimental conditions that it was not trained with. 

\end{NewBox}

\subsubsection*{STEP 4.1}
Assess the goodness of the fit achieved by the parameter optimisation procedure.

If the fit is not good, further action is required. Proceed to STEP 4.2.

\subsubsection*{STEP 4.2}
If the fit is not good enough, check convergence of the optimisation methods.
\begin{enumerate}
    \item If there are hints that searches were stopped prematurely (e.g. error messages that indicate that local optimisations did not converge), go back to STEP 3: modify the settings of the optimisation algorithms (e.g. increase maximum allowed time and/or number of evaluations) and run the optimisations again.
    \item If there are no signs of a premature stop, the problem may be that the optimal solution lies outside the initially chosen parameter bounds $\rightarrow$ go back to STEP 3: set larger parameter bounds and run the optimisations again.
    \item If the actions above do not solve the issue, it may be because the optimisation method is not well suited for the problem $\rightarrow$ go back to STEP 3: choose a different method and run the optimisations again.
\end{enumerate}

If the new optimisations performed in STEP 4.2 do not yet yield a good fit, there may be a problem with the choice of objective function. Proceed to STEP 4.3.

\subsubsection*{STEP 4.3}
If the fit is not good enough, go back to STEP 2 and select a different objective function.

If the new optimisation results are still inappropriate, the problem might be the model structure. Proceed to STEP 4.4.

\subsubsection*{STEP 4.4}
If the fit is not good enough, go back to the model equations and perform a model refinement.

\subsection*{STEP 5: Practical identifiability analysis}

\begin{figure*}
\centering
\includegraphics[width=0.9\linewidth]{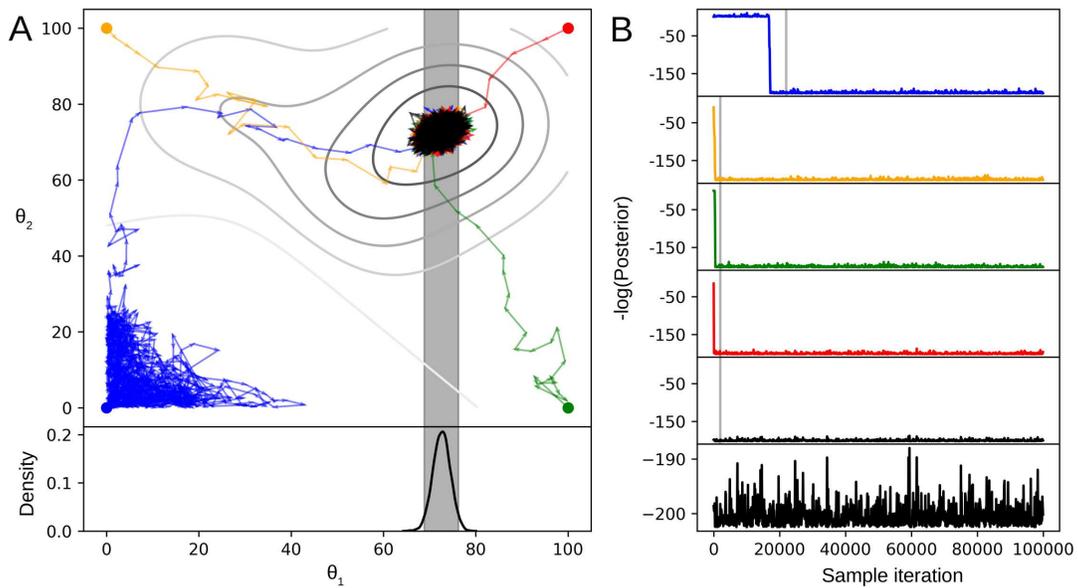}
\caption{\textbf{MCMC sampling.}
(A)
Upper: traces of MCMC chains through parameter space. The initial sample of a chain is indicated with ``$\bullet$''. Parameters are labelled as $\uptheta$ with an index as the subscript. The initial sample of the black chain is the maximum likelihood estimate (MLE) from an optimisation (at approximately $\uptheta_1 = \uptheta_2 = \textrm{70}$). Colour is used in all subfigures to indicate results corresponding to the same MCMC chain.
Lower: the marginal distribution and 95\% credibility interval for a parameter, given the black MCMC chain  without burn-in.
(B) Traces of the objective function value across the MCMC chains, including burn-in (indicated with vertical grey lines) as detected by the Geweke test. The bottom plot is a zoom-in of the second-to-bottom plot.
}
\label{fig:mcmc}
\end{figure*}

\begin{NewBox}*

\textbf{Box 5. Methods for STEP 5: Practical identifiability analysis}\\
\setlength\parindent{12pt}
The Fisher information matrix (FIM) is a widely used measure of the information content of the experimental data that provides information about the practical identifiability of the parameters. For a set of $n_t$ measurements it can be calculated as
\begin{equation}\label{FIM}
\sum_{j=1}^{n_y}
\sum_{i=1}^{n_t}\left(\frac{\partial{y_j}(t_i)}{\partial \theta}\right) W^{(i)} \left(\frac{\partial {y_j}(t_i)}{\partial \theta}\right)^T
\end{equation}
\noindent where $\frac{\partial{y}(t_i)}{\partial \theta}$ are the sensitivity functions and $ W^{(i)}$ is a diagonal matrix with $W_{jj}^{(i)} = 1/\sigma_j^2(t_i)$, where $\sigma_j(t_i)$ is the standard deviation.
The Cram\'er-Rao theorem \cite{cramer2016mathematical} states that, if $\hat{\theta}$ is an unbiased estimate of $\theta$ (i.e. $E(\hat{\theta}) = \bar{\theta}$), the inverse of the FIM is a lower bound estimate of the covariance matrix,
\begin{equation}
\text{Cov}(\hat{\theta}) \geq \text{FIM}^{-1}(\hat{\theta})
\end{equation}
The covariance matrix provides information about variability of individual parameters and of pairs across different realizations of the experimental data. It is defined as:
\begin{equation}
\text{Cov} = E\left[\left(\hat{\theta} - \bar{\theta} \right) \left(\hat{\theta} - \bar{\theta} \right)^T \right]
\begin{bmatrix}
\sigma^2(\hat{\theta}_1)             & \cdots & \text{cov}(\hat{\theta}_1 \hat{\theta}_{n_{\theta}})  \\
\vdots                          & \ddots & \vdots                           \\
\text{cov}(\hat{\theta}_{n_{\theta}} \hat{\theta}_1) & \cdots & \sigma^2(\hat{\theta}_{n_{\theta}}) 
\end{bmatrix}
\end{equation}
The chi-squared values follow an approximately Gaussian distribution \cite{Geier2012}. Confidence intervals estimated from the FIM are always symmetric and can be overly optimistic if nonlinearities are present, since they rely on linearisation of the models \cite{wieland2021structural}.
A more realistic -- albeit computationally more expensive -- alternative is to use profile likelihoods or sampling-based procedures \cite{banga-balsa:08}. 
The latter generate a large number of pseudo-experimental datasets and use them to solve different realizations of the parameter estimation problem. The resulting cloud of solutions is then used for estimating the confidence intervals and other statistical information. Several variants of this approach have been proposed in the literature, sometimes under the name ``bootstrap''. In a classic definition of bootstrap \cite{efron1986bootstrap}, different datasets ($S_1$, $S_2$, ...) are obtained by randomly sampling with replacement the original dataset $S$. In \cite{joshi-seidel-morgenstern-kremling:2006} the model is first calibrated with the original (true) experimental data $S$, yielding a parameter estimate $\hat{\theta}$. Subsequent datasets ($S_1$, $S_2$, ...) are generated by simulating the model with $\hat{\theta}$ and adding different realizations of artificial noise. In \cite{banga-balsa:08} it is suggested to obtain different parameters by solving the same parameter estimation problem (i.e. with the same dataset) starting from different initial points; this is similar to the ``multi-start'' procedure in \cite{frohlich2014uncertainty}. 
Another common resampling technique is the jackknife \cite{tukey1958bias,efron1981jackknife}, in which every sample is removed from the dataset once.

Another possibility is to use Bayesian sampling based procedures, which view parameters as random variables with a known prior distribution. Experimental data is used to compute a posterior distribution that describes the uncertainty of the problem \cite{toni2009approximate,liepe2014framework,hug2013high}. Since the prior distribution on the parameters is typically not available, it has been suggested to use the profile likelihood approach to estimate priors \cite{vanlier2012integrated}. Bayesian sampling methodologies, such as Markov chain Monte Carlo (MCMC), are often computationally expensive for large models. MCMC sampling is illustrated in Fig. \ref{fig:mcmc}.

The uncertainty quantification approaches mentioned so far can only be applied to structurally identifiable parameters, since they produce misleading results for structurally non-identifiable ones \cite{frohlich2014uncertainty} (recall that structural identifiability analysis should be performed before practical identifiability analysis). In contrast, the Profile Likelihood approach (PL) can be applied to structurally non-identifiable parameters, which are revealed by flat profiles.

\end{NewBox}

The task of quantifying the uncertainty in parameter estimates is known as practical (or numerical) identifiability analysis. It involves calculating univariate confidence intervals or multivariate confidence regions for the parameter values. Key concepts and tools for practical identifiability analysis are listed in Box 5. Practical identifiability issues are illustrated in Figures \ref{fig:bruno}D and \ref{fig:fujita}D.

\subsubsection*{STEP 5.1}
Perform practical identifiability analysis with one of the methods described in Box 5. If large uncertainties in parameter estimates are revealed, then proceed to STEP 5.2.

\subsubsection*{STEP 5.2}
If there are large uncertainties, then:
\begin{enumerate}
    \item If it is possible to perform new experiments $\rightarrow$
add more experimental data. In this case, the experiment should be optimally designed in order to yield maximally informative data. This is described in the following section.
    \item If it is not possible to perform new experiments $\rightarrow$ 
assess the possibility of simplifying the model parameterisation without losing biological interpretability.
    \item If neither (1) nor (2) are possible $\rightarrow$ 
include prior knowledge about parameter values. Such information (either about the value of a parameter or about its bounds) can sometimes be found in publicly available databases.
\end{enumerate}
After performing one of the above actions, go back to STEP 3.

\subsection*{(OPTIONAL STEP): Alternative experimental design for parameter estimation}

If practical identifiability analysis concludes that there are large uncertainties in the parameter estimates, a solution may be to collect new data. Ideally, it should be obtained by designing and performing new experiments in an optimal way. 
Optimal Experiment Design (OED) seeks to maximise the information content of the new experiments. It can be performed using optimisation techniques that minimise an objective function that represents some measure of the uncertainty in the parameters.
It is also possible to perform OED for other goals, such as model discrimination or decreasing prediction uncertainty.
OED techniques are discussed in Box (O).

\begin{NewBox}*

\textbf{Box (O). Optimal experimental design.}\\
\setlength\parindent{12pt}
Alternative experimental designs may increase the information content of the data, thereby improving the parameter estimates. If the best possible experiment is found via optimisation, the procedure is called optimal experimental design (OED). OED is formulated as a dynamic optimisation problem, in which the variables that can be changed are the experimental conditions (allowed perturbations, measured quantities, number of experiments, experiment duration, number and location of sampling times), and the objective to maximise is some measure of the information content of the data. The optimisation constraints are the system dynamics and the experimental limitations. The optimisation problem can be solved by control vector parameterization \cite{balsa-alonso-banga:2008}. 

OED can be performed with several purposes: decreasing the uncertainty in parameter estimates \cite{steiert2012experimental,banga-balsa:08,bock2013parameter,franceschini2008model,pronzato2008optimal}, decreasing the uncertainty in model predictions \cite{kreutz2013profile,hagen2013convergence}, improving controller performance \cite{gevers2005identification}, or discriminating between model alternatives \cite{casey2007optimal,waldron2019closed}. The objective function to optimise depends on this final goal. For the purpose of decreasing parameter uncertainty it is common to use objective functions based on the FIM (\ref{FIM}). Typical choices are the D-optimum and E-optimum criteria, which maximise the determinant of the FIM and its minimum eigenvalue, respectively. The D criterion minimises the geometric mean of the errors in the parameters, while the E-criterion minimises the largest error.

\end{NewBox}

\subsubsection*{STEP O.1}
Define the constraints of the new experimental setup, and, in case of optimal design, the criterion to optimise.

\subsubsection*{STEP O.2}
Obtain a new set of experiments, either by optimisation or from an educated guess.

\subsubsection*{STEP O.3}
Perform experiments and collect data.

\subsubsection*{STEP O.4}
Include the new data in the objective function and repeat STEPS 2--5.

\subsection*{STEP 6: Prediction uncertainty quantification}

If the calibrated model is used for making predictions, for example about the time course of its states, it is useful to assess the prediction uncertainty. This assessment is not trivial because uncertainty in parameters does not directly translate to uncertainty in predictions. Hence it is pertinent to quantify to which extent the uncertainty in model parameters leads to uncertainty in the predictions of state trajectories. 
Note that, if some parameters were fixed in STEP 1 to achieve structural identifiability, in this step their values have to be altered across the plausible regime to obtain realistic confidence intervals of the state predictions.
The available methods for prediction uncertainty quantification are reviewed in Box 6. Their application to case studies is shown in Fig. \ref{fig:bruno}E and Fig. \ref{fig:fujita}E.

\begin{NewBox}*

\textbf{Box 6. Methods for STEP 6: prediction uncertainty quantification}\\
\setlength\parindent{12pt}
Several techniques for quantifying the uncertainty of model predictions are available \cite{villaverde2019comparison}.

If the FIM (\ref{FIM}) is invertible, it is possible to approximate the uncertainty in the state trajectories by error propagation from the parameter estimates \cite{Geier2012}, with the caveats mentioned in Box 5, as $$\text{Cov}[x(t)]= \frac{\partial x(t,\theta)}{\partial\theta} \text{Cov}(\theta) \frac{\partial x(t,\theta)}{\partial\theta}^T$$
If the FIM is not invertible, as is the case if there are non-identifiable parameters, this approach cannot be directly applied. An alternative is to approximate the inverse of the FIM with the Moore-Penrose pseudoinverse \cite{shahmohammadi2019sequential}.

The prediction profile likelihood method (PPL) calculates confidence intervals for the states by performing constrained optimisation of the likelihood using a fixed prediction value as a nonlinear constraint \cite{kreutz2012likelihood}. It has been extended to calculate prediction bands via integration techniques \cite{hass2015fast}. Implementations of the PPL are available in the toolboxes Data2Dynamics \cite{raue2015data2dynamics}, PESTO \cite{stapor2017pesto}, and pyPESTO \cite{yannik_schalte_2021_4738579}.

The dispersion in model predictions can be quantified from an ensemble of calibrated models (ENS). Brown et al. used statistical mechanics considerations \cite{brown2004statistical} to build the ensemble. The consensus among ensemble predictions can be used to estimate the confidence in said predictions \cite{villaverde2015consensus}.

The possibility of adopting a Bayesian framework for quantifying the uncertainty in parameters was mentioned in Box 5. Accordingly, simulating the model with the sampled parameter vectors yields a sample from the prediction posterior (PP), thus allowing to assess prediction uncertainty \cite{vanlier2012integrated}.

A recent comparison of methods for prediction uncertainty quantification \cite{villaverde2019comparison} has found a trade-off between computational scalability and accuracy. The least computationally expensive method is the one based on the FIM, but it is also the least reliable. The method with worst scalability is the PP, which hampers its applicability to large models. PPL and ENS are more generally applicable than PP, and also more accurate than FIM.

\end{NewBox}

\subsubsection*{STEP 6.1}
Calculate confidence intervals for the time courses of the predicted quantities of interest using one of the methods in Box 6.

\subsection*{(OPTIONAL STEP): Model selection}

The protocol presented so far assumes that the model structure is known, except for the specific values of the parameters. Sometimes the form of the dynamic equations that define the model -- and not only the parameter values -- is not completely known a priori, and a family of candidate models may be considered. Model selection techniques choose the best model from the set of possible ones, aiming at a balance between model complexity and goodness of fit. They are discussed in Box (MS).

\begin{NewBox}*

\textbf{Box (MS). Model selection}\\
\setlength\parindent{12pt}
A simple way of comparing models is to see if the quality of their fits differ in a statistically significant way. This is known as a likelihood-ratio test.
A model with more parameters is more flexible, and it is therefore easier for it to achieve a better fit. However, an overparameterised model can exhibit overfitting, which is undesirable. To take this into account, when selecting a model one should aim at a balance between model complexity and goodness of fit. Measures such as the Akaike Information Criterion (AIC) \cite{bozdogan1987model} and the Bayesian Information Criterion (BIC) \cite{vyshemirsky2008bayesian} take into account the goodness of fit and a penalty based on the number of estimated parameters, and can thus be used to quantify this trade-off.

The trade-off between model complexity and goodness of fit can already be taken into account during parameter optimisation (STEP 3), by adding a sparsity-enforcing penalty in the objective function (STEP 2). In this way, the obtained parameter values correspond to a solution that represents an optimal trade-off. The weight given to the penalty controls the balance between both criteria. Increasing the weight of the penalization decreases the variance in the parameters at the expense of increasing their bias, an effect called shrinkage. The least absolute shrinkage and selection operator (LASSO) was introduced in \cite{tibshirani1996regression}. If the $L_1$ norm is used in the penalty, this approach is known as $L_1$-regularisation. A recent example of its application to dynamical biological models can be found in \cite{steiert20161}.

If it is feasible to perform new experiments, they may be specifically designed for the purpose of model discrimination, applying an experimental design procedure that seeks to maximise the difference between the outputs of candidate models (see Box (O)) \cite{waldron2019closed}.

Several different model structures may yield the same output, in which case they are called indistinguishable (similarly to a parameter being called non-identifiable if it can have an infinite number of values that lead to the same model output). When it is not possible to discriminate between the candidate models, a possibility is to take all of them into account. This can be done by building an ensemble of models, as described in Box 6, which contains not only models with different parameter vectors, but also different structures.

\end{NewBox}

\section*{Troubleshooting}

Troubleshooting advice can be found in Table \ref{Tab:troubleshooting}.

\begin{table*}[!ht]
\footnotesize%
\centering
\begin{tabular}{p{0.6cm} p{3.0cm} p{4.4cm} p{5.9cm}}
\toprule 
Step & Problem & Possible reason & Solution\\
\midrule
1    & It is not feasible to analyse structural identifiability due to computational limitations & The model is too large and/or too complex  & (A) Reduce the model complexity by fixing several parameters (conservative approach) (B) Use a numerical method (e.g. PL) to analyse practical identifiability as a proxy of structural identifiability \\
3    & Parameter optimisation takes very long & The size of the model makes this step computationally very expensive & Use parallel optimisation approaches to decrease computation times, or try a different optimiser\\
4     & Parameter optimisation does not result in a good fit & (A) The optimiser was stuck in a local minimum & (A) Use a global method and allow for enough time to reach the global optimum \\
     &  & (B) The parameter bounds are too small & (B) Set larger bounds \\
     &  & (C) The model is not an adequate representation of the system & (C) Modify the model structure \\
     &  &  &  In general: use hierarchical optimisation if applicable \\
4    & Parameter optimisation resulted in overfitting & Fitting the noise rather than the signal: very good calibration result that however generalises poorly & Use cross-validation to detect overfitting. If present: (A) Use regularisation in the calibration; (B) Simplify overparameterised models\\
5    & The confidence intervals of the parameters are very large & The data are not sufficiently informative to constrain the values of the parameters sufficiently & {(A) Add prior knowledge about parameter values and repeat the optimisation (B) Obtain new experimental data (ideally through OED) and repeat the optimisation} \\
6    & The confidence intervals of the predictions are very large & The data are not sufficiently informative to constrain the values of the predictions sufficiently & (Same as the above solution) \\
\midrule
\end{tabular}
\caption{\label{Tab:troubleshooting}Troubleshooting table. Common problems that may appear at different stages of the procedure, their causes, and solutions.} 
\end{table*}

\section*{Examples}

\subsection*{Carotenoid pathway model}

Our first case study is the carotenoid pathway model by Bruno et al. \cite{bruno2016enzymatic}, with 7 states, 13 parameters, and no inputs. The model output differs among the experimental conditions: in each of the six experimental conditions for which data is available, only one of the 7 state variables is measured (one is measured in two experiments, and two states are never measured). 

\begin{figure*}
\centering
\includegraphics[width=0.8\linewidth]{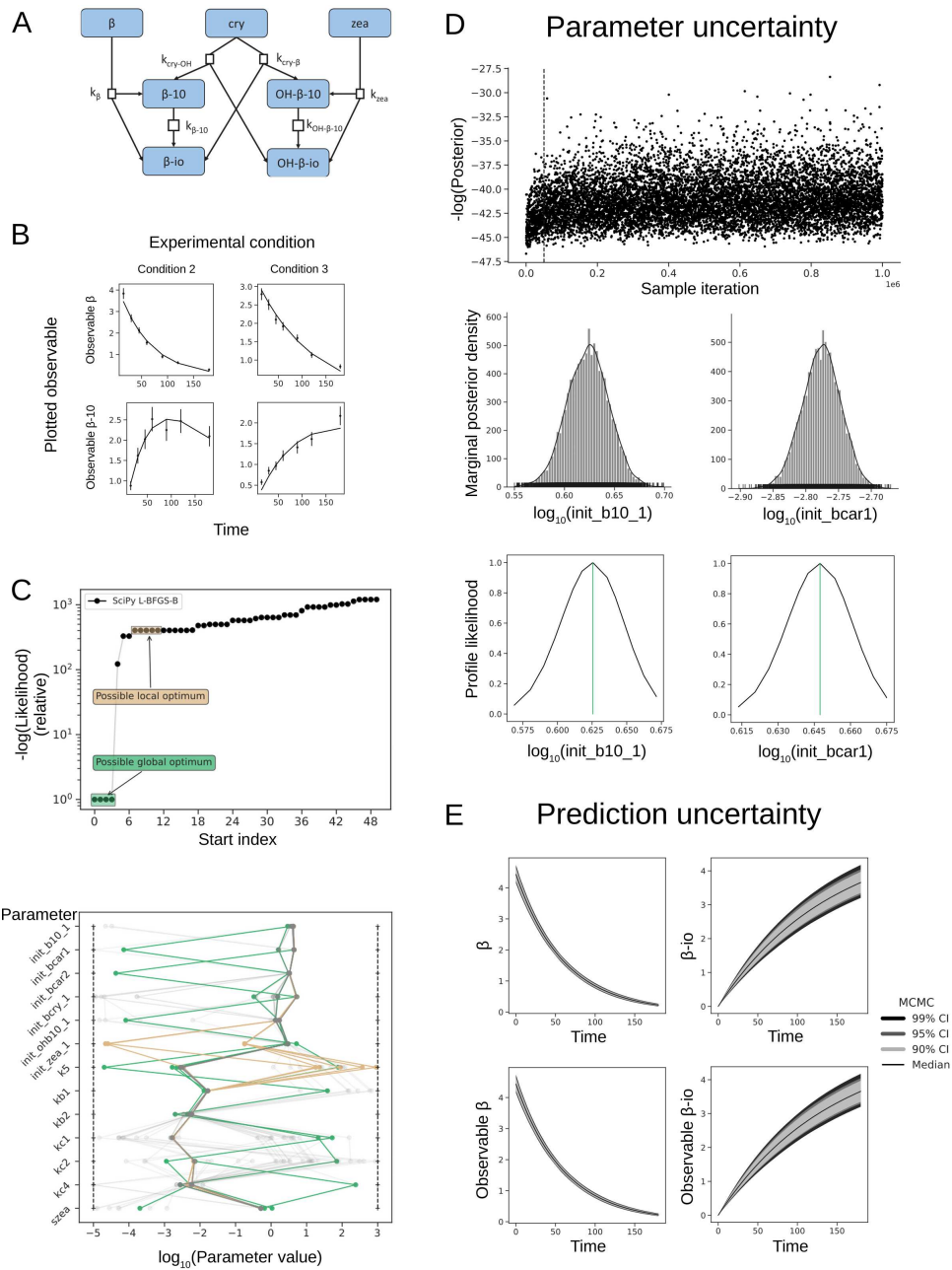}
\caption{
\textbf{Calibration of the carotenoid pathway model}.
(A) Schematic of the model pathway.
(B)  Visualization of the fit. The plot shows the trajectories of the model observables, as well as the means (points) and standard errors of the means (error bars) of the measurements.
(C)
Upper: A waterfall plot, showing the number of starts that converged to the MLE. Here and in the remaining subfigures, green indicates results that correspond to the MLE.
Lower: A parameters plot, showing variability of parameters among starts that converged to the possible global optimum (green). Vertical dotted lines indicate parameter bounds.
(D) Plots related to parameter uncertainty analysis.
Upper: a trace of the function values of samples from a MCMC chain. The vertical dotted line indicates burn-in.
Middle: marginal density distributions of two parameters, using samples from the converged chain. The plots show a kernel density estimate, histogram, and rug plot.
Lower: profiled likelihood of two parameters.
(E) Plots related to prediction uncertainty analysis, computed as percentiles from predictions of samples.
Upper: prediction uncertainties of two states.
Lower: prediction uncertainties of two observables.
Note that in this model, observables are states without transformation, hence the observables and states have the same uncertainties.
}
\label{fig:bruno}
\end{figure*}

The application of the protocol is summarised in the following paragraphs, and the main results are shown in Fig. \ref{fig:bruno}.

\subsubsection*{STEP 1.1: Structural identifiability analysis}
We first assess structural identifiability and observability for each individual experimental condition, obtaining a different subset of identifiable parameters for each one. Next, we repeat the analysis after combining the information from all experiments, obtaining that all parameters are structurally identifiable. However, the two state variables that are not measured in any experiment ($\upbeta$-io and OH-$\upbeta$-io) are not observable. If the initial conditions of these two states were considered as unknown parameters, they would be non-identifiable.

\subsubsection*{STEP 1.2: Address structural non-identifiabilities}
We are not interested in the two unobservable states. Hence we omit this step, and proceed with the original model.

\subsubsection*{STEP 2.1: Objective function}
We use the negative log-likelihood objective function described in Equation \ref{eq:llk}, which is the common choice in frequentist approaches.

\subsubsection*{STEP 3.1 and 3.2: Parameter optimisation}
We estimate model parameters using the multi-start local optimisation method L-BFGS-B implemented in the Python package SciPy. With 100 starting points we achieve convergence to the maximum likelihood estimate, as indicated in the waterfall plot (Fig. \ref{fig:bruno}). The parameters plot shows that the parameter vector is similar amongst the best starts, indicating that the parameters are well-determined by the optimisation problem and the optimiser.

\subsubsection*{STEP 4.1: Assess goodness of fit}
Visual inspection indicates a good quality of the fit, with simulations closely matching measurements.

\subsubsection*{STEP 4.2: Address fit issues}
As the fit is good, this step is skipped.

\subsubsection*{STEP 5.1: Practical identifiability analysis}
We analyse practical identifiability using profile likelihoods and MCMC sampling. Profile likelihoods suggest that all parameters are practically identifiable, as the confidence intervals span relatively small regions of the parameter space. The profiles peak at the maximum likelihood estimate (MLE), suggesting that optimisation was successful.
MCMC sampling yields similar results; parameter marginal distributions span a similar distance of parameter space compared to profile likelihoods, and credibility intervals are also similar.

\subsubsection*{STEP 6.1: Prediction uncertainty analysis}
We calculate credibility intervals using ensembles of parameters from sampling. In this model, there is a one-to-one correspondence between states and observables, hence the plots are the same. The prediction uncertainties are reasonably low, suggesting that the model has been successfully calibrated and might be used to predict new behaviour.

\subsection*{Akt pathway model}

\begin{figure*}
\centering
\includegraphics[width=0.8\linewidth]{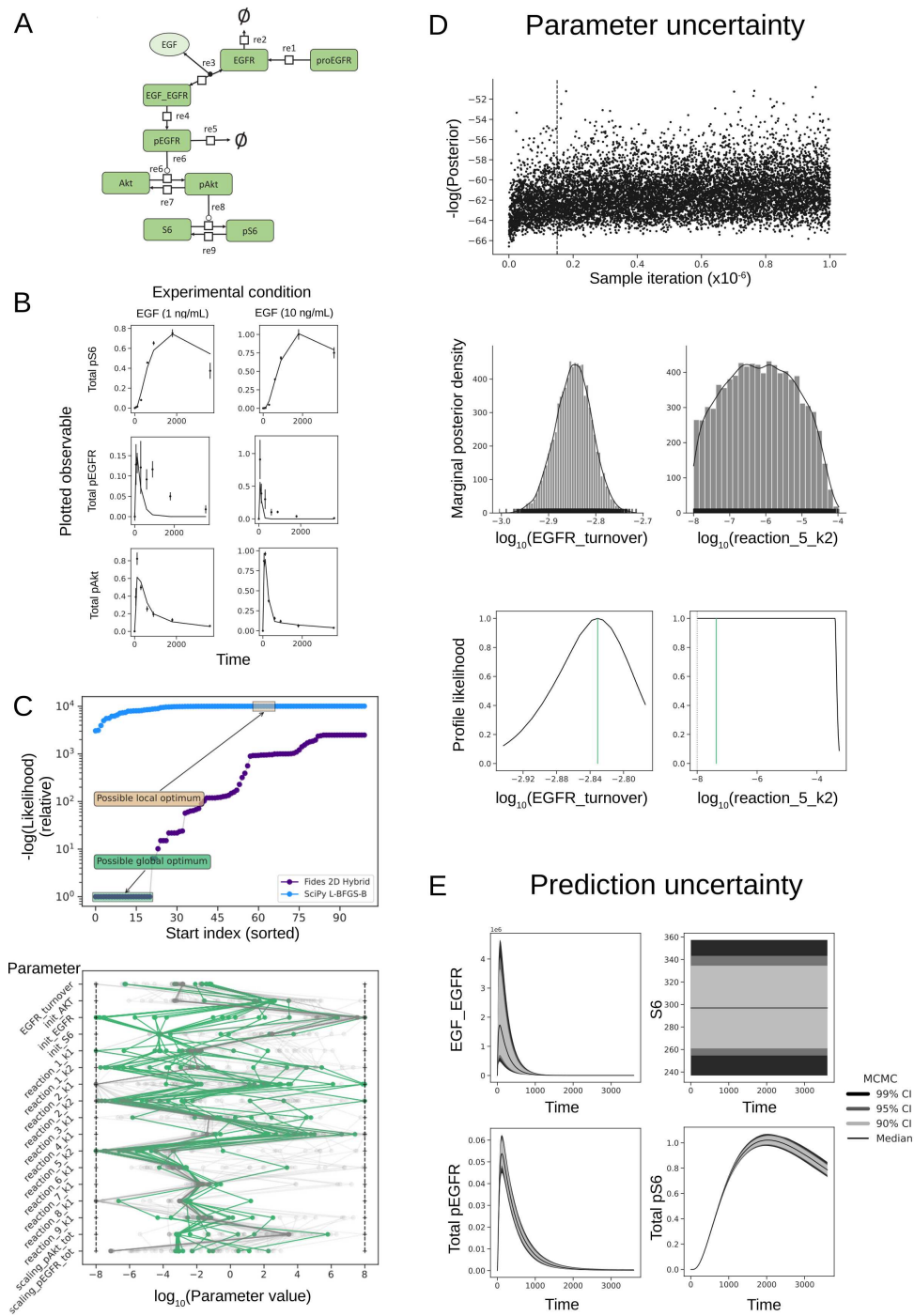}
\caption{
\textbf{Calibration of the Akt pathway model}.
(A) Schematic of the model pathway.
(B) Visualization of the fit. The plot shows the trajectories of the model observables, as well as the means (points) and standard errors of the means (error bars) of the measurements.
(C)
Upper: A waterfall plot, showing the number of starts that converged to the MLE. Here and in the remaining subfigures, green indicates results that correspond to the MLE.
Lower: A parameters plot, showing variability of parameters among starts that converged to the possible global optimum (green). Vertical dotted lines indicate parameter bounds.
(D) Plots related to parameter uncertainty analysis.
Upper: a trace of the function values of samples from an MCMC chain. The vertical dotted line indicates burn-in.
Middle: marginal density distributions for two parameters, using samples from the converged chain. The plots show a kernel density estimate, histogram, and rug plot.
Lower: profiled likelihood of two parameters. The dotted vertical line indicates a parameter bound.
(E) Plots related to prediction uncertainty analysis, computed as percentiles from predictions of samples.
Upper: prediction uncertainties of two states under one experimental condition.
Lower: prediction uncertainties of two observables under one experimental condition.
}
\label{fig:fujita}
\end{figure*}

The second example is an AKT pathway model \cite{fujita2010decoupling} with 22 unknown parameters, 3 of which are unknown initial conditions, 9 state variables, 3 outputs, and 1 input. There are 6 experimental conditions, each of them with a different input EGF concentration. 

Results are summarised in the following paragraphs and in Fig. \ref{fig:fujita}.

\subsubsection*{STEP 1.1: Structural identifiability analysis}
 We consider the following scenarios:
\begin{enumerate}
    \item For a single experiment with constant EGF, 11 parameters are structurally non-identifiable, and 3 states are unobservable.
    \item For a single experiment with time-varying EGF, the model becomes structurally identifiable and observable. 
    \item For multiple experiments (at least two) with constant EGF, the model is structurally identifiable and observable. 
\end{enumerate}
The experimental data available corresponds to the scenario (3) above. The scenario (2) yields an identifiable and observable model, but it requires a continuously varying value of EGF, which is not practical. 
It is also interesting to note the role of initial conditions in this case study. The results summarised above are obtained with generic (nonzero) initial conditions. However, in the available experimental datasets there are several initial conditions equal to zero. Introducing this assumption in the analyses of the scenarios (2) and (3) leads to a loss of identifiability and observability: four parameters become non-identifiable and one state becomes unobservable.

\subsubsection*{STEP 1.2: Address structural non-identifiabilities}
We assume a realistic scenario corresponding to the available experimental data: several experimental conditions with a constant input, EGF, and certain initial conditions equal to zero. In this case the model has four non-identifiable parameters and one unobservable state. To make the model fully observable and structurally identifiable, it is necessary and sufficient to fix the value of two of the non-identifiable parameters. 
Thus, we fix two of these parameters and proceed with the next steps. 

For comparison, we also performed the remaining steps without fixing the non-identifiable parameters. We found that fixing the non-identifiability issues resulted in slightly faster and more robustly convergent optimisations, as well as in better practical identifiability and reduced state uncertainty.

\subsubsection*{STEP 2.1: Objective function}
We choose the negative log-likelihood objective function described in Equation \ref{eq:llk}.

\subsubsection*{STEP 3.1 and 3.2: Parameter optimisation}
Similarly to the other case study, we initially use the multi-start local optimisation method ``L-BFGS-B''. 

\subsubsection*{STEP 4.1: Assess goodness of fit}
Visual inspection (i.e. comparison of the simulations produced by the maximum likelihood estimate with the measurements) reveals a poor fit to the data (not shown). This result is obtained even with the best result obtained from thousands of optimisation runs from different starting points. 

\subsubsection*{STEP 4.2: Address fit issues}
First we try to improve the fit by tuning the settings of the optimisation method, L-BFGS-B, without success. 
Then we try a different method, Fides, which has a higher computational cost but achieves higher quality steps during optimisation. With Fides we find an MLE that produces a fit comparable to the one reported in the original publication. The high number of starts (in the order of $10^3$) required to find this fit reproducibly indicates that this is a difficult parameter optimisation problem.

\subsubsection*{STEP 5.1: Practical identifiability analysis}
Credibility intervals obtained from MCMC sampling indicate that several parameters are practically non-identifiable. This result is not significantly improved by fixing parameters as suggested in STEP 1.2. Improving the practical identifiability of these parameters would require repeating the calibration with additional experimental data.

\subsubsection*{STEP 6.1: Prediction uncertainty analysis}
Credibility intervals obtained from MCMC sampling indicate that the uncertainties in the observable trajectories are reasonably low. However, the state trajectories have larger uncertainties, which make this calibrated model unsuitable for predictions involving these states. The quality of the predictions can be improved by reducing practical non-identifiabilities in the model, as mentioned in the previous step.

\section*{Discussion and conclusion}

In this paper we have proposed a pipeline of methods and resources for calibrating ODE models in the context of biological applications. Its end goal is to obtain a model that is capable of making predictions about quantities of interest with quantifiable uncertainty. 

The pipeline consists of a series of steps, each of which represents a task that should be fulfilled before proceeding to the next one to ensure a successful calibration. Performing these tasks entails applying computational methods of different types, symbolic and numerical. The analyses and calculations can be computationally challenging in practice. While the protocol is not dependent on a particular choice of software, we have recommended a number of state-of-the-art tools that implement the methods. 

To facilitate the application of the protocol by novices as well as by experienced modellers, we have described in detail how to perform each of the protocol steps. We have also provided the theoretical background required for understanding the underlying problems. Furthermore, we have illustrated its use with two case studies: a carotenoid pathway model in \textit{Arabidopsis thaliana}, and an EGF-dependent Akt pathway of the PC12 cell line.
Finally, we have highlighted some of the most common pitfalls in biological modelling, showing how to avoid them.

\section*{Key Points}
\begin{itemize}
\item The correct calibration of dynamic models is essential for obtaining correct predictions and insights. 
\item While a wide range of tools and resources are currently available, there are also many potential pitfalls, even for the expert. 
\item Here we propose a model calibration protocol that covers all aspects of the problem.
\item The present paper guides the user through all the steps of the pipeline, providing a one-stop guide that is at the same time compact and comprehensive.
\item We provide all the code required to reproduce the results and perform the same analysis on new models, so that the biological modelling community can benefit from this pipeline.
\end{itemize}

\section*{Supplementary data}

All data, scripts, and examples presented in this paper can be downloaded from:\\ \url{https://github.com/ICB-DCM/model_calibration_protocol_preprint}. 


\section*{Funding}

This project has received funding from the European Union's Horizon 2020 research and innovation programme under grant agreement No 686282 (``CANPATHPRO''). JRB also acknowledges funding from the Spanish MINECO/FEDER project SYNBIOCONTROL (DPI2017-82896-C2-2-R). AFV was partially supported by a Ram\'on y Cajal Fellowship (RYC-2019-027537-I) from the Spanish Ministry of Science, Innovation and Universities.
Funding was received from the Deutsche Forschungsgemeinschaft (DFG, German Research Foundation) under Germany's Excellence Strategy (EXC 2151 - 390873048: JH; EXC-2047/1 - 390685813: DP), and the German Federal Ministry of Economic Affairs and Energy (Grant no. 16KN074236: DP).


\end{document}